\documentclass[11pt,twoside]{article}
\usepackage{./asp2014}

\aspSuppressVolSlug
\resetcounters
\newcommand{\HI}{\mbox{H\,{\sc i}}}
\newcommand{\HII}{\mbox{H\,{\sc ii}}}

\newcommand{\kms}{\mbox{km~s$^{-1}$}}
\newcommand{\juc}{\mbox{$J$=1$-$0}}
\def\lsim{~\rlap{$<$}{\lower 1.0ex\hbox{$\sim$}}}
\def\gsim{~\rlap{$>$}{\lower 1.0ex\hbox{$\sim$}}}
\newcommand{\jdu}{\mbox{$J$=2$-$1}}
\newcommand{\doceCO}{\rm $^{12}$CO}
\newcommand{\treceCO}{\rm $^{13}$CO}
\newcommand{\jtd}{\mbox{$J$=3$-$2}}
\begin{document}

\title{Evolved Stars}
\author{Lynn D. Matthews$^1$ and Mark J Claussen$^2$
\affil{$^1$Massachusetts Institute of Technology Haystack Observatory,
  Westford, MA, 01886 USA; \email{lmatthew@mit.edu}}
\affil{$^2$National Radio Astronomy Observatory, P.O. Box O, Socorro,
  NM, 87801 USA; \email{mclaussen@nrao.edu}}}

% This section is for ADS Processing.  There must be one line per author.
\paperauthor{Lynn D. Matthews}{lmatthew@mit.edu}{}{Massachusetts
  Institute of Technology}{Haystack Observatory}{Westford}{MA}{01886}{USA}
\paperauthor{Mark
  J Claussen}{mclaussen@nrao.edu}{ORCID_Or_Blank}{National Radio
  Astronomy Observatory}{}{Socorro}{NM}{87801}{USA}

%Chapters do not have abstracts
\begin{abstract}
This chapter reviews some of the expected contributions of 
the ngVLA to the understanding of the late evolutionary stages 
of low-to-intermediate mass stars, including asymptotic giant 
branch (AGB) stars, post-AGB stars, and pre-planetary nebulae. 
Such objects represent the ultimate fate of stars like the Sun, 
and the stellar matter they lose to their immediate vicinity 
contributes significantly to the chemical enrichment of galaxies. 
Topics addressed in this chapter include continuum imaging of 
radio photospheres, studies of circumstellar envelopes in both 
thermal and nonthermal lines, and the investigation of the 
transition stages from the AGB to planetary nebulae using radio 
wavelength diagnostics. The authors gratefully acknowledge 
contributions to the content of this chapter from members of 
the evolved star community.

\end{abstract}

\section{Introduction}
%
%To fill in this template, make sure that you read and follow the ASPCS
%Instructions for Authors and Editors available for download online.
%\footnote{Most URLs should be in a footnote like this one.  In this
%  case, you can download the online material from
%  \url{http://www.aspbooks.org}.}  
%Further hints and tips for including graphics, tables, citations, and
%other formatting helps are available there, in addition to the
%examples given in the aspauthor.tex file included here.  

Evolution onto the asymptotic giant branch (AGB) marks the final
thermonuclear burning stage
in the lives of low-to-intermediate mass stars ($0.8 \lsim M_{*} \lsim
8~M_{\odot}$).
The AGB is characterized by a dramatic increase in
luminosity (factors of $10^{4}$ or more) and significant mass-loss
($\dot{M}\sim10^{-8}$ to $10^{-4}M_{\odot}$~yr$^{-1}$) through
cool, low-velocity ($\sim$10~\kms) winds.
Over the course of roughly one million years,
this mass loss leads to the formation of extensive
circumstellar envelopes (CSEs) of gas and dust that may span more than
a parsec in size.  

Knowledge of the mass-loss histories of AGB stars is fundamental
to understanding
the complete life-cycle of our Galaxy. The winds of AGB stars are rich
in heavy elements and molecules, sowing the seeds for future
generations of star and planet formation. Indeed,
mass loss during the AGB is believed to be responsible for $\sim$50\% of the chemical
enrichment in the Galaxy (e.g., Van Eck et al. 2001). However, many
aspects of the lives of AGB stars remain poorly understood, including
the wind driving mechanisms,  the geometry of AGB mass loss,  and the
evolutionary paths for stars of different initial masses
(see H\"ofner \& Olofsson 2018).

Radio observations play a crucial role in understanding stellar
evolution on the AGB and beyond. Continuum
radiation is emitted from the optically thick
radio photospheres of AGB stars, as well as from
the jets and torii that may appear during subsequent evolution to the post-AGB and
pre-planetary nebula (pPN) stages. 
The cool, extended
atmospheres and CSEs of  evolved stars also
give rise to numerous spectral lines at cm and mm wavelengths, offering powerful
diagnostics of the chemistry, temperature, and density structure of the CSE, as
well as 
wind outflow speeds and other kinematic information. Studies of radio
lines supply crucial insight into the stellar mass-loss
history and probe how evolved, mass-losing stars interact with their interstellar
environments and shape the composition and small-scale structure of
the ISM. The specific properties of
CSEs gleaned from radio observations also supply insight into their
role in the subsequent
evolution into planetary nebulae (PNe) and Type~Ia supernovae.

The ngVLA will provide enormous gains in
our understanding of the end stages of the lives of Sun-like stars. Its exquisite
sensitivity will permit imaging observations of large numbers of
spectral lines with high angular and spectral resolution, yielding the most
detailed and comprehensive
maps to date of CSEs from the smallest to the largest scales. The
ultrawide bandwidths will make it possible to observe  simultaneously both lines and
the (relatively weak) stellar continuum emission,
providing precise astrometric registration required for the testing of
sophisticated radiative transfer and hydrodynamical
modeling of AGB star atmospheres and envelopes.

%%%%%%%%%%%%%%%%%%%%%%%%%%%%%%%%%%%%%%%%%%%%%%%%%%%%%%
 \begin{figure}[h]
   \vspace{-0.0cm}
   \hspace{3cm}
     \rotatebox{0}{
\resizebox{!}{6cm}
{\includegraphics{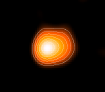}
     }}
         \caption{The radio photosphere of  R~Leo imaged with the
           VLA at 7~mm (Matthews et al. 2018). The star shows a non-spherical shape
           and evidence for a non-uniform surface. The image was produced using a sparse model
           reconstruction algorithm, enabling
           super-resolution of $\sim$0.75 times the dirty beam
           FWHM of $\sim$38~mas. The peak flux density is $\sim$4~mJy
           beam$^{-1}$ and the image is $\sim$150~mas on a side. The
           ngVLA will enable comparable resolution on AGB stars at
           distances of up to 1~kpc as well as the resolution of finer
           surface details on nearby stars. }
    \label{fig:photosphere}
\end{figure}
%%%%%%%%%%%%%%%%%%%%%%%%%%%%%%%%%%%%%%%%%%%%%%%%%%%%%%%%%

\section{Continuum Observations of Radio Photospheres\protect\label{photospheres}}
AGB stars emit optically thick continuum radiation at cm and
(sub)mm wavelengths from a region at $\sim2R_{\star}$ known as the
``radio photosphere'' (Reid \& Menten 1997; see also Harper, this volume).  In a typical
oxygen-rich AGB star,  the radio
photosphere has a diameter of $\sim$4~AU and lies near the outskirts of the so-called molecular
layer or ``MOLsphere''
(Tsuji 2000, 2001), but interior to the dust-formation zone and
maser-emitting regions (see Section~\ref{masers}). Thus the  radio
photosphere samples an atmospheric region 
impacted by  pulsation, convection, shocks, and other key processes
responsible for the transport of material from the stellar surface to the outflowing wind.

Presently it is possible to resolve the radio surfaces of the nearest
AGB stars ($d\lsim$150~pc) at $\lambda\lsim$1~cm using the VLA and ALMA in their most
extended configurations.  Such observations have led to tantalizing
evidence for non-uniform radio surfaces and non-spherical photospheric shapes that evolve over time, most
likely as a result of pulsation and convective effects (Reid \& Menten
2007; Matthews et
al. 2015b, 2018; Vlemmings et al. 2017; Figure~\ref{fig:photosphere}).   
The ngVLA is expected to provide an angular resolution of $\sim$26~mas at 8~GHz
and $\sim$2~mas at 93~GHz\footnote{Throughout this chapter we assume a
  maximum baseline length of $\sim$300~km (Selina \& Murphy 2017).};
thus at shorter wavelengths, 
the ngVLA will be able to resolve the
surfaces of AGB stars to distances beyond a kpc, making it possible to
assess the ubiquity of 
spots and other surface features and measure their brightness temperatures and
temporal
evolution.  For nearby stars ($d\lsim$150~pc), 
it will be possible to
resolve radio photospheres over nearly two decades in frequency,
effectively providing ``tomography'' of the atmosphere (see Lim et
al. 1998; O'Gorman et al. 2015; Matthews et al. 2015b).
The ngVLA will be complementary to ALMA for such work,
as cm bands probe higher layers of the atmosphere than
can be studied with ALMA and also suffer negligible contamination from
line emission and dust. The further addition of contemporaneous imaging with optical/infrared interferometers such as
the Center for High Angular Resolution Astronomy (CHARA) Array and
Very Large Telescope Interferometer (VLTI) would
enable the most stringent tests to date of sophisticated 3D models of AGB
star atmospheres  (e.g., Freytag et al. 2017).

Sensitivities of $\sim0.4\mu$Jy beam$^{-1}$ in one hour at
$\lambda\approx$7~mm 
will also permit the 5$\sigma$ detection of unresolved radio
photospheres to distances of at least 5~kpc. Over the course of
roughly a year, the  radio light curve observations of Reid
\& Menten (1997) found levels of variability in a small sample
of AGB stars at 8~GHz to be $\lsim$15\%,
implying that shock speeds within the radio photosphere are limited to $\lsim$5~\kms. But at 7~mm,
evidence is now seen for larger flux density changes in data
taken several years apart (Matthews et al. 2015b, 2018). The ngVLA will
enable for the first time 
studies of the variability of radio photospheres over both short
(intra-pulsation cycle) and long (several years) timescales for a
statistical samples of stars. Such data will be extremely
powerful for understanding the roles of shocks in governing the
physics of AGB star atmospheres, as well as for gathering for the
first time 
statistics on the frequency of possible eruptive or flaring activity
in AGB stars. Evidence for the latter phenomena has already been seen at other
wavelengths, including the ultraviolet (Sahai et al. 2011b, 2015)
and in X-rays (Karovska et al. 2005). Such events may be triggered by accretion
activity linked with a companion, chromospheric-like activity, or other
still unknown mechanisms.

\section{Spectral Line Studies of the CSEs of  AGB Stars}
\subsection{Molecular Lines (Thermal)}
The rich chemistry and cool temperatures of the extended atmospheres
and CSEs of AGB stars give rise to emission from
a multitude of molecular lines at cm and mm wavelengths, predominantly
through rotational transitions (e.g., Turner \& Ziurys 1988). Together
these various line
transitions provide important diagnostic information about the range
of temperature and density of the CSE and its kinematics. However,
with the exception of masing lines (see Section~\ref{masers}), the bulk of these lines are
quite weak, requiring excellent sensitivity for their
identification and mapping, particularly at the high angular
resolutions needed to spatially resolve their distribution and kinematics within the CSE.

\subsubsection{Astrochemistry and Searches for New Circumstellar Molecules}
The leap in sensitivity of the ngVLA is expected to lead to the
discovery of numerous additional molecular species in the envelopes of AGB stars.
These are likely to include pre-biotic molecules, carbon chains, and
other complex molecules suspected of having their
origin in AGB star winds (e.g., Ziurys 2006). The frequency
range covered by the ngVLA will have advantages over ALMA for
searching for many complex molecules in CSEs, both because heavier molecules
have lower frequency rotational transitions, and because line blending
below $\sim$100~GHz is much less severe than at higher frequencies. 
Even for commonly observed molecular transitions such as the 
CO $J=1-0$ line, the ngVLA will provide important new statistics by enabling
detection of late-type stars to distances of up to $\sim$10~kpc,
including within the Galactic Center Region.

\subsubsection{Spiral Patterns in the Mass Ejecta of AGB Stars}
Binary motion can generate spiral patterns in the CSEs of
AGB stars, and modeling of the spiral pattern can be used to set constraints on
orbital parameters 
(e.g., period, eccentricity, and inclination; see Mastrodemos \&
Morris 1999; Kim \& Taam 2012a, b, c). 
Interferometric molecular-line imaging of CO lines has detected such
patterns in a small number of objects 
(e.g., Maercker et al. 2012, Kim et al. 2017). For the carbon star
CIT~6, the spiral 
pattern was found via VLA imaging of the 
HC$_3$N(4-3) at 36.3~GHz (Claussen et al. 2011). However, the unprecedented
sensitivity of ngVLA 
will enable searches for such patterns in a statistical sample of AGB stars.

\subsection{Non-Thermal Lines (Masers)\protect\label{masers}}
Molecular maser emission is frequently detected in the
CSEs of AGB stars (and in young PNe; see below).   The most
widely observed masers in circumstellar environments include
transitions of OH (1.6~GHz), H$_{2}$O (22~GHz),  SiO $v$=0,1,2,3,
\juc\ (43~GHz), and
SiO $v$=1, \jdu\ (86~GHz) (e.g., Alcolea 2004).  These various transitions require
different combinations of temperature and density to excite, hence
observations of multiple transitions within a single star can be
used to probe material from different regions in the CSE, ranging
from a few AU from the star for SiO masers to $r\sim$1000~AU for OH lines.

\begin{figure}[h]
%   \vspace{-1.cm}
   \hspace{0.75cm}
     \rotatebox{0}{
\resizebox{!}{10.5cm}
{\includegraphics{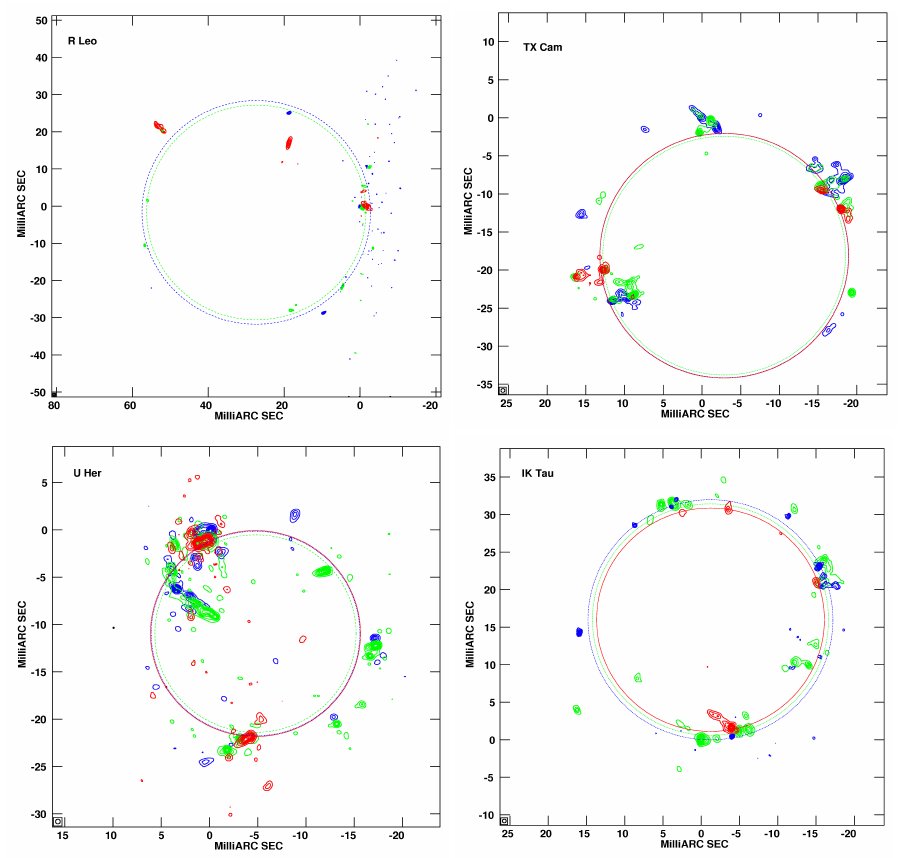}
     }}
         \caption{Very Long Baseline Array (VLBA) maps of SiO $v$ = 1
           (in blue), $v$=2 
           (in green), and $v$=3 (in red), \juc\ maser lines from 
           R~Leo, TX~Cam, 
U~Her, and IK~Tau (Desmurs et al. 2014).
           The relative positions of the maser spots of the different
           lines should provide clues to understanding the SiO pumping, but
           presently the relative astrometry is very uncertain.}
    \label{fig:masers}
\end{figure}

The strongest circumstellar masers show
brightness temperatures of many millions of K, and the
intensity and high excitation of some of these lines have enabled 
exquisitely high resolution mapping of CSEs using Very Long Baseline Interferometry
(VLBI; e.g.,  Cotton et al. 2004; Gonidakis et
al. 2013; Desmurs et al. 2014; Figure~\ref{fig:masers}).  However,
there are various limitations of VLBI studies, including losses of
up to $\sim$50\% of the
flux density (e.g., Desmurs et al. 2017) and
the difficulty of phase  referencing (and thus, astrometry) at 
frequencies above $\sim$20~GHz (Beasley \& Conway 1995).  The ngVLA will be able to systematically observe
circumstellar masers with excellent angular resolutions
($\sim$10~mas at 22~GHz). But in contrast to current VLBI arrays, it will 
have much higher sensitivity, negligible flux loss, and excellent
instantaneous $u$-$v$-coverage. The high signal-to-noise ratios (SNR)
that will be achieved will additionally enable highly
accurate positioning of maser lines (in principle, the achievable
astrometric precision is
approximately the angular resolution divided by the SNR).
And crucially, the ultra-wide bandwidths of the 
ngVLA will permit the routine simultaneous detection of multiple maser
transitions, along with the
stellar photosphere in the continuum, allowing  placement of
the distributions of the different masers in the context of the inner circumstellar
layers.   In nearby stars, it will also be possible to study in detail
the movements of maser spots
and the relation between the structure of different maser species
during the course of the stellar pulsation.
These types of data will be vital for identifying  the maser pumping
mechanisms and providing quantitative constraints on models of AGB
star atmospheres (e.g, Gray et al. 2009, 2016).

%%%%%%%%%%%%%%%%%%%%%%%%%%%%%%%%%%%%%%%%%%%   
%%%%%%%%%%%%%%%%%%%%%%%%%%%%%%%%%%%%%%%%%%%%%%%%%%%%%%%   

The ngVLA performance will also enable systematic high-resolution mapping of
relatively weak maser lines  ($T_{b}\lsim10^4$~K), whose properties are  not yet well understood and for
which very little observational information  is currently available. Examples include
the SiO $v$=0 masers  (useful to study intermediate  shells of
CSEs; Bolboltz \& Claussen  2004); $^{29}$SiO $v$=0,1, $^{28}$SiO $v$=2,
\jdu, and $^{28}$SiO $v$=3 masers (which are important for the
study of SiO maser pumping; Soria-Ruiz et al. 2005, Desmurs et
al. 2014), and  HCN \juc\  masers  at 88.6~GHz (which may be the  best way  to
probe the very inner  regions in C-rich CSEs; Lucas et  al. 1988, Menten
et al. 2018). 

The study of circumstellar
OH masers will benefit from the ngVLA's excellent angular resolution even
at longer wavelengths (18~cm), along with a factor of several gain in line sensitivity over the
present VLA. This will make OH masers around late-type stars accessible throughout the 
Galaxy and enable new constraints on brightness temperature and maser
gain models.  
In addition to surveys, follow-up imaging studies
with a combination of high sensitivity, high angular resolution, high spectral resolution 
($\le$0.07~km s$^{-1}$), and full Stokes parameters will enable
separation of  
individual maser components and detection of the often present signatures of strong linear 
or circular polarization that are otherwise washed out by insufficient spatial
and velocity 
resolution (e.g., Wolak et al. 2012, 2013). 

A special category of OH maser sources is the
OH/IR stars, a subclass of evolved stars believed to be in the
final stages of their evolution on the AGB. Their dusty winds and exceptionally high
mass-loss rates (up to a few times $10^{-4}~M_{\odot}$ yr$^{-1}$)  render them optically thick at visible
wavelengths, and they 
generally emit a strong characteristic ``doubled-peaked'' 1612~MHz OH
maser emission profile (e.g., Engels 1985).
To test the predictions of evolutionary models, in particular
the role of metallicity, observational properties of OH/IR
stars during their obscured phase {\em and} reliable distances
are needed. For OH/IR stars, the latter can be derived via the so-called
``phase-lag'' method (Engels et al. 2015;
Etoka et al. 2017), using a combination of: (i)
the linear diameter of the shell as obtained from the light curve
of the varying OH maser peaks; and (ii) an angular diameter
obtained from an interferometric map of the OH emission.
In order to get a precise angular diameter determination
it is vital to map the faint ``inter-peak'' emission from which
the full extent of the shell and its geometry can be inferred.
While challenging with current instruments, the enhanced sensitivity and resolution of the ngVLA
will facilitate such measurements for larger samples of OH/IR stars. For
the nearest stars ($d\lsim$200~pc) it should become possible to do direct
comparisons  of such distances with OH maser parallax measurements.

\subsection{The \HI\ 21-cm Line\protect\label{HI}}
\begin{figure}[h]
 %  \vspace{-0.5cm}
   \hspace{2.cm}
     \rotatebox{0}{
\resizebox{!}{12cm}
{\includegraphics[width=3in]{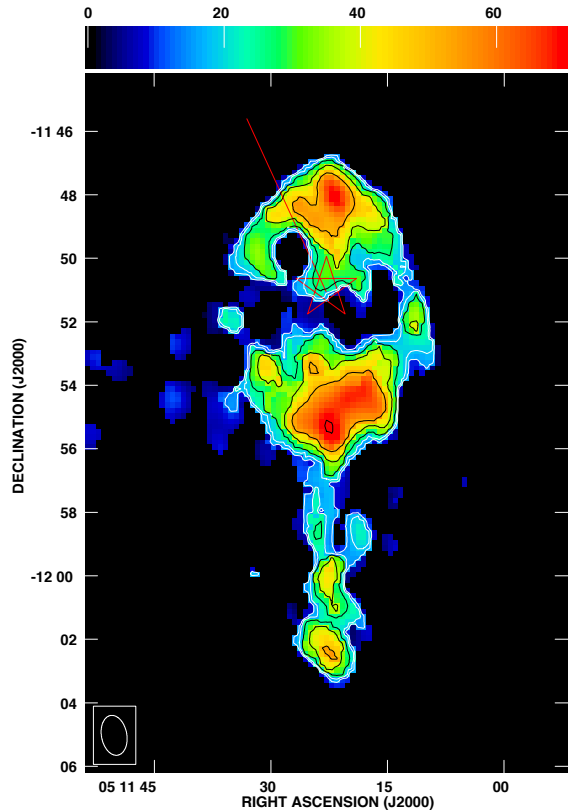}
     }}
         \caption{VLA \HI\ total intensity image of the AGB star RX~Lep (Matthews
  et al. 2013). The intensity scale units are Jy beam$^{-1}$ m
  s$^{-1}$. 
A star symbol denotes the stellar position and the
  direction of space motion is shown by a red line. The projected
  extent of the
  \HI\ shell and tail in the north-south direction is $\sim$0.7~pc.
 }
    \label{fig:HI}

\end{figure}

The CSEs surrounding AGB stars can become enormously extended, in some
cases spanning more than a
parsec (e.g., Villaver et al. 2012). As
densities drop with increasing distance from the star,  molecules, 
including CO, become dissociated by the interstellar
radiation field. Molecular lines therefore cannot probe the
outer regions of the CSE. 
However, neutral atomic hydrogen (\HI)
persists in this environment, and recent studies have demonstrated
that its 21~cm line emission provides a
powerful probe of outer CSE (representing $\gsim10^{5}$ yr of
mass loss history), as well as the interfaces between AGB
stars and their environments (e.g., Figure~\ref{fig:HI}).  While
FIR and/or far-ultraviolet emission can sometimes supply complementary
information on the CSE outskirts, only \HI\
studies supply {\em kinematic} information on the extended gas (e.g., Matthews
et al. 2008).

%%%%%%%%%%%%%%%%%
%%%%%%%%%%%%%%%%%%%%

Surveys carried out with the Nan\c{c}ay Radio
Telescope have now identified dozens of candidate \HI\ shells around
nearby AGB
stars (G\'erard \& Le~Bertre 2006; G\'erard et al. 2011 and in prep.). However, follow-up interferometric observations 
are critical for studying the detailed spatial and kinematic structure of the \HI\
envelopes and for disentangling
possible contamination from line-of-sight
emission. Only a handful of AGB stars have been imaged in the \HI\ line
with the VLA to date, owing to the challenging and time-consuming nature of these observations 
(e.g., Matthews \& Reid 2007; Matthews et al. 2008, 2013). However,
the ngVLA will provide substantial gains in sensitivity, survey speed, and in the 
mitigation of 
background confusion for circumstellar \HI\ observations.

The quasi-random  distribution of the
individual antennas in the ngVLA core  will 
considerably improve the $u$-$v$ coverage compared with the existing
VLA and reduce the periodic sidelobe effects caused by regularly
spaced antennas, thereby helping to mitigate the problems of
contamination from interstellar \HI\ emission along the line-of-sight.
(cf. Matthews et al. 2011, 2013). The addition of a large
($\gsim$45~m) single-dish antenna
to the array core (Frayer 2017) would also significantly enhance the ability to
subtract off line-of-sight emission and to image extremely low surface
brightness \HI\ emission from \HI\ shells, tails, and
other structures spanning tens of arcminutes (see e.g.,
Matthews et al. 2013, 2015a). 

Even in the
nearest AGB stars, circumstellar \HI\
emission is typically weak with narrow linewidths ($\sim$10~\kms), requiring good
spectral resolution ($\lsim$1~\kms)  to measure velocity gradients and
other kinematic features (e.g., Matthews et al. 2008, 2011, 2013). 
For sightlines or velocity ranges 
that are not confusion limited, the sensitivity gained by the factor of $\sim$4 increase in collecting
area of the ngVLA will enable expansion of the
study of circumstellar \HI\ to significantly larger samples of stars, 
Finally, the ngVLA core configuration is expected to have baselines up to 1.55~km,
which will provide $\sim$50\% higher spatial resolution than the current VLA
D configuration for mapping the complex kinematics of the emission,
including cases where the \HI\ emission may trace the extension of
axisymmetric outflow structures seen at smaller radii in molecular lines (Hoai et al. 2014).

\section{Evolution Past the AGB}
\subsection{Understanding the AGB$\rightarrow$PN Transition}
The mass loss characteristic of evolved stars
increases with evolutionary stage (Balick \& Frank 2002,
Bujarrabal et al.\ 2001).
During the pPN
stage, the star ejects a significant fraction of 
its initial mass ($\sim 1~M_{\odot}$) via a ``superwind''
within only a few thousand years (e.g., Lewis 2001). During this phase, the 
core of the star becomes visible and illuminates (and
will later ionize) a very massive nebula, forming a PN, whose central
stellar core will
then evolve into a white dwarf (WD). However, very few details are known
about how these events unfold.  

Observations of molecular lines at cm and mm wavelengths are a powerful
tool for studying the AGB/PN transition, including the disks, torii, and
outflows that commonly appear during this stage---as well as the
central stars themselves (e.g., Alcolea et
al. 2007; Bujarrabal et al. 2013; Sahai et al. 2017). The ngVLA will complement 
studies performed with ALMA by extending the range of molecular
transitions that can be observed with exquisite sensitivity and
spatial resolution to
include such lines as 
\doceCO\ \&
\treceCO\ \juc\ (110-115~GHz), HCN \juc\ (89~GHz), SiO $v$=1,
\juc\ \& \jdu\ (43 and 87~GHz),  CS $J=1-0$ \& $2-1$ (49 and 98~GHz). At these frequencies, the ngVLA is
expected to achieve 
angular resolutions of $\sim$2--5~mas---of order ten times
better than present ALMA resolutions.  At the same time, 
the ngVLA's continuum sensitivity will allow
probing the dust 
component of the central torii in a statistical sample of post-AGB
objects, determine their cm-mm spectral energy distributions, and
constrain the grain sizes and dust masses. Combined, this information will
address key questions about the geometry and origin of these
structures (e.g., Bondi-Hoyle accretion vs. 
common-envelope ejection).

Even at very high spatial and spectral resolutions,
the impressive sensitivity of the ngVLA will allow maps of lines
arising from hot inner regions, where
brightness temperatures of several hundred K are expected. The innermost
layers of pPNe also develop very compact \HII\ regions, resulting from
the evolution of the central star to a WD and the ensuing ionization of
the surrounding material. Therefore, those regions trace the last phases of
the nebular ejection. Their continuum emission is opaque or nearly
opaque and very intense at most ngVLA frequencies ($T_{b}$>1000~K);
recombination lines with $T_{b}$ $\sim$10\% of the continuum (i.e., $>$100~K) are also expected
(e.g., Kwok \& Bignell 1984; S\'anchez Contreras et al. 2017).  The
ngVLA will therefore enable studies  with high accuracy of both the
neutral and the ionized gas from very inner regions of pPNe, where
vital clues 
to understand the late evolutionary phases of stars are most likely
hidden.

%%%%%%%%%%%%%%%%%%%%%%%%%%%%%%%%%%%%%%%%%%%%%%%%%%%%%%%
\begin{figure}
%   \vspace{-1.cm}
\center{\fbox{%
\includegraphics[width=5.in,angle=0]{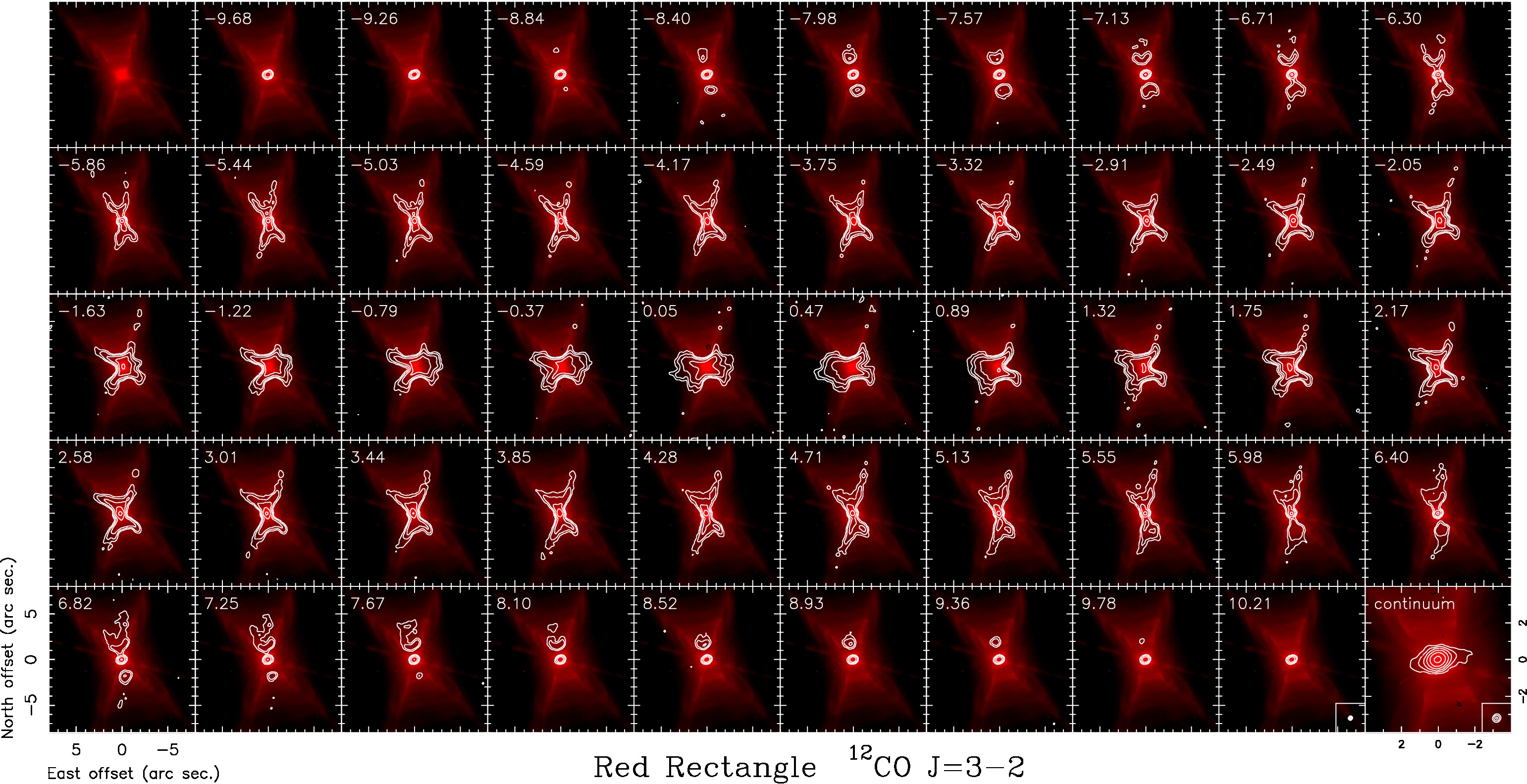}}}
\caption{ALMA observations of \doceCO\ \jtd\ emission from the pPN object
           the Red Rectangle (Bujarrabal et
           al. 2013; contours). A {\it Hubble Space Telescope} optical
           image is also shown in color.  The
           equatorial feature at the central velocities is a rotating
           disk; the {\sf X}-shaped features come from more diffuse gas in
           expansion.  It is
           thought that rotating disks in the centers of pPNe can
           account for their characteristic
           bipolar shapes, but only very high angular resolution allows
           mapping the relevant components.}
\label{fig:redrec}
\end{figure}
%%%%%%%%%%%%%%%%%%%%%%%%%%%%%%%%%%%%%%%%%%%%%%%%%%%%%% 

\subsection{The Role of Binarity in Post-AGB Evolution}
Binary companions are thought to play a significant role in the
late evolutionary stages of Sun-like stars.  
For example, binary interactions are widely believed to underlie the formation of most 
PNe and may hold the key to the resolution of a
long-standing puzzle, namely that although PNe evolve from AGB stars,
whose CSEs typically appear spherical with relatively, slow, isotropic
expansion
($V_{\rm exp}\sim$\,5-15\,\kms), the vast 
majority of PNe and pPNe exhibit axisymmetric structures, with a variety of elliptical, 
bipolar, and multipolar morphologies, as well as fast, collimated outflows
($V_{\rm exp}\gsim$\,50-100\,\kms; e.g., Figure~\ref{fig:redrec}). 

About half of post-AGB objects are known or likely
binaries and show prominant disks with
little or no extended nebulosity. These
have been dubbed ``dpAGB'' objects, and their
dusty disks are remarkably similar to those of pPNe in
harboring substantial 
masses of very large (mm-sized) dust grains (e.g., Sahai et
al. 2011a). This
supports the idea that the formation of dense ``waists'' 
in post-AGB objects is intimately linked to binarity. Surveys of
dpAGBs and pPNe with the ngVLA to measure the cm-wavelength SEDs of a
large population of these targets will allow the characterization of
the dust disks in both classes of objects and new insights into the role of binarity.

\subsection{Water Fountain Sources}
Water fountain sources (WFs) are thought to be one of the first manifestations of highly
collimated outflows in the very short transition from the AGB to PNe (see, e.g.,
Claussen et al. 2007).  Simultaneous ngVLA observations
of the continuum and maser emission from WFs in full polarization mode
at frequencies of 1.6~GHz (OH masers), 22~GHz (H$_{2}$O masers), and 43~GHz (SiO
masers)  are expected to significantly advance our
understanding of these sources and hence this important but
short-lived evolutionary stage of Sun-like stars. Observations with
the ngVLA of optically thin thermal lines
such as HCN, 
H$^{13}$CN $J=1-0$, CS $J=1-0$ \& $2-1$, and SO $3,2-2,1$ are also
expected to provide unique insights (Sahai et al. 2017).

Studies of cm wavelength continuum from WFs 
will enable the detection and study of the expected thermal radio jets and their
powering sources. Even though thermal free-free emission produced by
photoionization of the gas in the parental CSE is not
expected to be significant [the central star(s) are too cold],
thermal free-free emission from shock-ionized gas could be present. A
similiar phenomenon is
observed in low-mass young stellar objects, where
ionization of the  gas in the local environment may occur through strong shocks of the wind
against the ambient gas (e.g., Rodr\'\i guez et al. 1999).
Radio jets have been detected towards the objects in the
pPNe stage, but it will be crucial to explore of the youngest
phase of these 
jets (i.e., during the WF phase) 
in order to fully understand their underlying driving mechanism.

From the formulation given in Torrelles et al. (1985) and
Curiel et al. (1989), it is  estimated that a wind with a
mass-loss rate of $10^{-7}~M_{\odot}$ yr$^{-1}$ and a velocity of
500~km s$^{-1}$ can produce via shocks
thermal free-free emission at a level of $\sim1~\mu$Jy at $\lambda\sim$1~cm
for a source at $d\approx$10~kpc. This emission can be detected with the ngVLA
with SNR$\approx$10 for an unresolved source (assuming rms noise 
0.1 $\mu$Jy beam$^{-1}$ after 10 hours of
integration). By studying the radio continuum emission at different
frequencies with the ngVLA, together with the kinematics of
masers in the outflow, it will be possible to obtain basic parameters of the central source of
the WFs (e.g., mass-loss rate and wind velocity) and to
understand the physical relationship between the masers and the
continuum (e.g., Orosz et al. 2018).

Observations of the maser emission from WFs with the ngVLA
over multiple epochs will also
enable measurements of proper motions and enable the derivation
of 3D space velocities. In particular, this
will be possible for 22~GHz H$_{2}$O masers, where
individual maser clumps have been observed to be moving at velocities of
$\sim$100~km s$^{-1}$ on time scales of months for sources at 10~kpc
distance. Because the continuum emission will be observed
simultaneously, it will be possible to derive relative positions of
the masers with respect to the central source with high precision. This is
mandatory to enable detailed modeling of the motions of
the gas, from which one may then extract information on the physical properties
of the envelope created during the AGB phase (e.g., its gas density
and mass; 
Orosz et al. 2018). As the mass-loss rate climbs to $\sim10^{-5}$ -- $10^{-4} M_{\odot}$ yr$^{-1}$
during the 
superwind phase, it will also be possible to test WF models by searching for
evidence of  deceleration of the jets using the spatio-kinematics of the
masers  (Imai et al. 2005).

Finally, in addition to allowing measurements of proper motions, observations 
of H$_{2}$O masers in WFs with the ngVLA will provide
new information on the relationship between the magnetic
field and velocity vectors  through
polarization measurements of the masers and their 3D
velocities. Similar techniques have already been applied to the
studies of YSOs (e.g., Goddi et al. 2017; see also
Hunter et al., this volume). This will be invaluable for
testing the role of magnetic fields in jet launching.

\acknowledgements The authors gratefully acknowledge contributions
to this Chapter from the following
individuals: Javier Alcolea; Valent\'in  Bujarrabal; Dieter Engels; Sandra Etoka;
Eric G\'erard; Jos\'e-Francisco Gomez; Hiroshi
Imai; Thibaut Le~Bertre; Anita M. S. Richards; Raghvendra Sahai; Jos\'e-Maria Torrelles.

%%%%%%%%%%%%%%%%%%%%%%%%%%%%%%%%%%%%%%%%%%
%\bibliography{editor}  % For BibTex
% For non-BibTex:

\end{document}